\documentclass[letterpaper]{article} 
\usepackage{aaai23}  
\usepackage{times}  
\usepackage{helvet}  
\usepackage{courier}  
\usepackage[hyphens]{url}  
\usepackage{graphicx} 
\usepackage{natbib}  
\usepackage{caption} 
\usepackage{amssymb}
\usepackage{amsmath}
\usepackage[T1]{fontenc}

\def \h {\mathbf{h}}
\def \c {\mathbf{c}}
\def \m {\mathbf{m}}
\usepackage{algorithm}
\usepackage{algorithmic}

\usepackage{newfloat}
\usepackage{listings}
\usepackage{multirow}
\usepackage{booktabs}
\usepackage{color}
\DeclareCaptionStyle{ruled}{labelfont=normalfont,labelsep=colon,strut=off} 
\floatstyle{ruled}
\newfloat{listing}{tb}{lst}{}
\floatname{listing}{Listing}
\newcounter{definition}[section]
\renewcommand{\thedefinition}{\arabic{definition}}
\newenvironment{definition}{
	\refstepcounter{definition}
	{\vspace{1ex} \noindent\bf  Definition \thedefinition:}}{
	 \vspace{1ex}} 

\let\oldAA\AA
\renewcommand{\AA}{\text{\normalfont\oldAA}}

\title{Predicting  Protein-Ligand Binding Affinity with Equivariant Line Graph  Network}
\author {
    Yiqiang Yi\textsuperscript{\rm 1},
    Xu Wan  \textsuperscript{\rm 1},
    Kangfei Zhao \textsuperscript{\rm 2},
    Le Ou-Yang \textsuperscript{\rm 1},
    Peilin Zhao \textsuperscript{\rm 2}
}
\affiliations {
    \textsuperscript{\rm 1} Shenzhen University, Shenzhen, China\\
    \textsuperscript{\rm 2} Tencent AI Lab, Shenzhen, China\\
}

\usepackage{bibentry}

\begin{document}

\maketitle

\begin{abstract}
\vspace{-0.1in}
Binding affinity prediction of three-dimensional (3D) protein ligand complexes is critical for drug repositioning and virtual drug screening.
Existing approaches transform a 3D protein-ligand complex to a two-dimensional (2D) graph, and then use graph neural networks (GNNs) to predict its binding affinity.
However,  the node and edge features of the 2D graph are extracted  based on invariant local coordinate systems of the 3D complex.
As a result, the method can not fully learn the global information of the complex, such as, the physical symmetry and the topological information of bonds.
To address these issues, we propose a novel Equivariant Line Graph Network (ELGN) for affinity prediction of 3D protein ligand complexes. 
The proposed ELGN firstly adds a super node to the 3D complex, and then builds a line graph based on the 3D complex.
After that, ELGN uses a new E(3)-equivariant network layer to pass the messages between nodes and edges based on the global coordinate system of the 3D complex.
Experimental results on two real datasets demonstrate the effectiveness of ELGN over several state-of-the-art baselines.
\end{abstract}

\section{1 \quad Introduction}
\vspace{-0.05in}
Computational drug discovery plays an important role in shorten the research cycle and reduce the cost and risk of failure in discovering novel drugs \cite{bahuguna2020overview,theodoris2021network}. In computational drug discovery, one of the most important tasks is to predict the binding affinity of a drug for the correct protein target, which is crucial for evaluating the efficacy of a drug. Unlike experimental measuring methods that utilize time-consuming biological tests to screen candidate molecules, machine learning-based  methods can rank drug candidates efficiently by predicting the binding affinities between ligands (i.e., drugs) and proteins \cite{pgal_kdd21}. 

The advances in protein structure prediction technologies promotes the accumulation of three-dimensional (3D) structure protein data  \cite{callaway2020will}. 3D structural information have proved useful for the design of drugs \cite{edm_icml22,pocket2mol_icml22,shan2022deep}, for example, recent studies learn the representations of protein-ligand complexes with atomic coordinates to predict their binding sites~\cite{graphbp_icml22,stark2022equibind}. Similar to binding site prediction, predicting the binding affinities between proteins and ligands need to learn the representations of protein-ligand complexes, which is still challenging to handle the coordinates.

Two typical types of methods have been proposed to learn the representations of protein-ligand complexes. The first type learns the representations of proteins and ligands separately, and then concatenate the representations together. For example, convolutional neural networks (CNNs) have been used to learn the representations of proteins and ligands based on the amino acid sequence data of proteins and the  SMILES of ligands \cite{ozturk2018deepdta,thafar2022affinity2vec,li2022bacpi}.  Graph neural networks (GNNs) have also been used to learn the representations of ligands \cite{nguyen2021graphdta}. However, they ignore the graph topological information of proteins, and  the interactions between proteins and ligands. 

The second type of methods learns the representations of protein-ligand complexes directly. Although CNN-based methods could learn the representations of protein-ligand complexes \cite{bty374,ragoza2017protein,zheng2019onionnet}, they are not able to capture the topological information. Moreover, most of the existing  methods ignore the spatial information of proteins and ligands \cite{yang2019analyzing,lim2019predicting,danel2020spatial}.
SIGN \cite{pgal_kdd21} employs a nearest neighbor technique to build an interaction graph. Then it utilizes GNNs architecture to learn the representation of the interaction graph and learn the spatial information (e.g angle) through polar coordinates. However, all these methods transform a 3D protein-ligand complex to a 2D graph, which  will lead to loss of the spatial information. And these methods also ignore the topological information of bonds. The topology of bonds is a new perspective in spatial information,  and has a positive effect on the learning of the complex embedding. Moreover, GNNs usually consisits of only two or three layers due to the over-smoothing phenomenon, so that each node can not aggregate messages from far atoms~\cite{gnnmp_iclr20,decgcn_neurips21}.


In this paper, we propose a novel deep learning framework, Equivariant Line Graph Network (ELGN) for protein-ligand binding affinity prediction. ELGN adopts E(3)-equivariant neural message passing layers to learn graph representation on the 3D protein-ligand interaction graph, where nodes denote the atoms of the ligand and protein and bonds are the spatial relationships among the atoms. 
In ELGN, we model the topology of bonds by a line graph and explicitly encode the spatial information of the bonds via learning on the line graph.  
We devise the architecture of atom-bond block to alternatively learn the atom embeddings on the interaction graph and the bond embeddings on the line graph, where the bond embeddings are used to enhancing the representation of the atoms as well as the whole graph.  
We add a super node that connects to all the atoms into the protein-ligand interaction graph to promote global feature fusion, and help each node to receive messages from far atoms. 
To further improve the generalization capability of ELGN, we employ a self-supervised learning task to predict the co-occurrent frequency of atom pairs in a larger region of ligand and protein interface by an atomic type pooling mechanism, which enabling the model to capture long-distance correlation of atoms.
The main contributions of this paper are summarized as follows: 
\begin{itemize}
\item We introduce an interaction line graph to model the topology of the bonds of the protein-ligand complex, which fully incorporates  modeling of the bonds into the learning procedure and enhances the learned representation of the complex.
\item We propose  Equivariant Line Graph Nework (ELGN) for protein-ligand binding affinity prediction. To capture 3D spatial coordinates of the complex, ELGN jointly learns the interactions of the atoms and interactions of the bonds by neural passing passing layers. 
\item Experiments on two real-world datasets verify that the proposed ELGN consistently outperforms 11 state-of-the-art methods on various metrics, which achieves a relative improvement of 4.2\% in RMSE on average.
\end{itemize}

\section{2 \quad Related Work}
\vspace{-0.05in}
\subsubsection{Protein-Ligand of Binding Affinity}
The task of protein-ligand binding affinity prediction in drug discovery is crucial, especially for drug screening. Recently, deep learning-based models have proliferated and excelled in performance.  DeepDTA first trains  CNNs to learn the representations independently from the amino acid sequences of proteins and the features of ligands, which are then combined to predict affinity\cite{ozturk2018deepdta}. However, this approach disregards information on protein-ligand interactions and spatial information. Because of the powerful capacity of GNNs to learn graph representations with topology information, GraphDTA leverages GNNs to learn the representation of a ligand \cite{nguyen2021graphdta}. The topology information of the ligand can be captured, but this information cannot model accurately the 3D spatial structure information. \cite{ragoza2017protein, stepniewska2018development}  leverage 3D convolutions (3D-CNNs) to capture spatial information. \cite{danel2020spatial,pgal_kdd21} inject distance and angle information into GNNs to learn the representations of protein and ligand. Due to these methods learning spatial information from 2D protein-ligand complex, one limitation is that the structures of proteins and ligands are invariant in complexes, such as rotation and translation, which implies that information about the spatial structure is unavoidably imperfect. In this paper, we learn spatial information on 3D protein-ligand complex with equivariance to predict protein-ligand binding affinity.   
\subsubsection{Equivariance} Equivariance  is a highly significant research in molecular representation. Intuitively, If we rotate or translate a ligand in 3D, the ligand embedded inside should also rotate or translate in the same way. Equivariance is crucial to the acquisition of our spatial knowledge. To this end, \cite{egnn_icml21} proposed the SE(3)-equivariant graph neural network (EGNN) that directly uses coordinates for message passing, while maintaining the equivariance of rotation and translation of molecules and proteins. And it effectively improves the ability to learn spatial information. The EGNN has numerous variations \cite{plgnn_aaai22,se3gnn_icml22}. Additionally, the equivariant has a positive impact on molecular linker design \cite{3dlinker_icml22} and molecular representation \cite{chiro_iclr22}. However, the 3D protein-ligand complex is typically converted into a 2D graph in protein-ligand affinity prediction in order to learn the representations of protein and ligand \cite{danel2020spatial,pgal_kdd21}. As a result, the 2D graph fails to consider the 3D spatial information, such as equivariance. To overcome this limitation, our model  directly learns the equivariant spatial informaton based on 3D protein-ligand complexes.

\subsubsection{Line Graph}
The topology of edges in graphs is neglected  and instead concentrate on the topology between nodes. However, edge topology is crucial for many tasks. The line graph is applied in bioinformatics, protein interaction networks \cite{pereira2004detection} and reaction network \cite{nacher2004clustering}. \cite{article} leveraged the line graph to uncover overlapping communities of its nodes. It is believed that the line graph offers an additional, and useful representation of the system's topology. Recently, \cite{hsu2022efficient} modelled the geometry of atomic structures based on line graph. \cite{liu2022graph} proposed Dual Message Passing Neural Networks based on the idea of line graph. \cite{dn4gl_icml22} proved that dummy nodes and the line graph can boost graph structure learning and alleviate the over-smoothing problem of GNNs. Nevertheless, the topological information of bonds is ignored in molecular representations \cite{nguyen2021graphdta,chiro_iclr22}. In our model, we utilize the line graph and graph convolutional network (GCN) to learn the topological information of bonds.

\section{3 \quad Preliminaries}
\vspace{-0.05in}
In this section, we introduce some definitions and notations used in our model and formulate this work.

A ligand is modeled as an undirected graph, $\mathcal{G}_l = (\mathcal{V}_l, \mathcal{E}_l, C_{l})$, where $\mathcal{V}_l$ is a set of atoms as the nodes, $\mathcal{E}_l$ is a set of bonds as the edges, $C_l \in \mathbb{R}^{|\mathcal{V}_l|\times 3}$ denotes the 3D coordinates of the $|\mathcal{V}_l|$ atoms.  For an atom $a_i$, we use $c_i$ to denote its coordinate. 
Given a ligand $\mathcal{G}_l$, we represent the protein-ligand complex as a protein-ligand interaction graph, which models the relationship of ligand atom and the protein atoms in the contact interface of the complex. 

\begin{definition} Protein-ligand Interaction Graph is a triplet $\mathcal{G} = (\mathcal{V}, \mathcal{E}, C)$.  Specifically,  $\mathcal{V} = \mathcal{V}_l \cup \mathcal{V}_p$ is the node set, containing the ligand atom set $\mathcal{V}_l$ and a set atoms $\mathcal{V}_p$ from a protein, and $C = [C_l, C_p]$ contains the coordinates of atoms in $\mathcal{V}_l$, $C_l$, and coordinates of atoms in $\mathcal{V}_p$, $C_p$, correspondingly. Here, $\mathcal{V}_p = \{ a_j |  \exists a_i \in \mathcal{V}_l,  \|\mathbf{c}_i - \mathbf{c}_j \| < d\}$ is a set of protein atoms  in the protein-ligand interface, within a cutoff distance $d$ between any atom $a_i$ in the ligand. 
The edges of $\mathcal{G}$ contains 3 types of bonds, i.e., $\mathcal{E} = {\mathcal{E}_l} \cup \mathcal{E}_{lp} \cup \mathcal{E}_p$, including the bonds of the ligand $\mathcal{E}_l$, the bonds between ligand atoms and protein atoms, $\mathcal{E}_{lp} = \{ (a_i, a_j) | a_i \in \mathcal{V}_l,  a_j \in \mathcal{V}_p, s.t. \|\mathbf{c}_i- \mathbf{c}_j\| < d\}$, and the bonds between protein atoms $\mathcal{E}_p = \{ (a_i, a_j) |, a_i, a_j \in \mathcal{V}_p, s.t. \|\mathbf{c}_i- \mathbf{c}_j\| < d \}$.
\end{definition}

Here, the cutoff distance $d$ restricts that only partial protein atoms that are close to the ligand are involved in the modeling, which determine the binding affinity to be predicted. The nodes (atoms) and edges (bonds) of the interaction graph have their features, denoted as $F_{\mathcal{V}} \in \mathbb{R}^{|\mathcal{V}| \times p}$ and $F_{\mathcal{E}} \in \mathbb{R}^{|\mathcal{E}| \times q}$, respectively. We use $F_{a_i} \in \mathbb{R}^{p}$ and $F_{b_{ij}}$ to represent the $p$-dim feature of an atom $a_i$ and the $q$-dim feature of a bond $b_{ij}$ in the following.
In this paper, to further model the interaction of the bonds in the interaction graph $\mathcal{G}$,  we construct the line graph~\cite{article} of $\mathcal{G}$ as defined below.

\begin{definition}
Interaction Line Graph. Given a protein-ligand interaction graph $\mathcal{G} = (\mathcal{V}, \mathcal{E}, C)$, the interaction line graph is a an undirected graph $\mathcal{G}' = (\mathcal{V}', \mathcal{E}')$, where $\mathcal{V}' = \{ b_{ij} | b_{ij} \in \mathcal{E}\}$, and $\mathcal{E}' = \{ (b_{ij}, b_{jk}) | b_{ij}, b_{jk} \in \mathcal{E} \}$. 
\end{definition}

Intuitively, the nodes of the line graph is the bonds of the interaction graph, the edges of the line graph is bond pairs if and only if two bonds of the pair connect to a same atom.
\textbf{Problem Statement.} In this work, we aim to building a neural network regression model $f_{\theta}(\mathcal{G}, \mathcal{G'}, F_{\mathcal{V}}, F_\mathcal{E})$ to predict the binding affinity of a protein-ligand complex, given the protein-ligand interaction graph $\mathcal{G}$, the interaction line graph $\mathcal{G}'$, the atom and bond features $F_{\mathcal{V}}$ and $F_{\mathcal{E}}$, where $\theta$ is the neural network parameters. The model $f_{\theta}$ should preserve equivariance regarding the atom coordinates $C$.

\begin{figure*}[]
\centering
\includegraphics[]{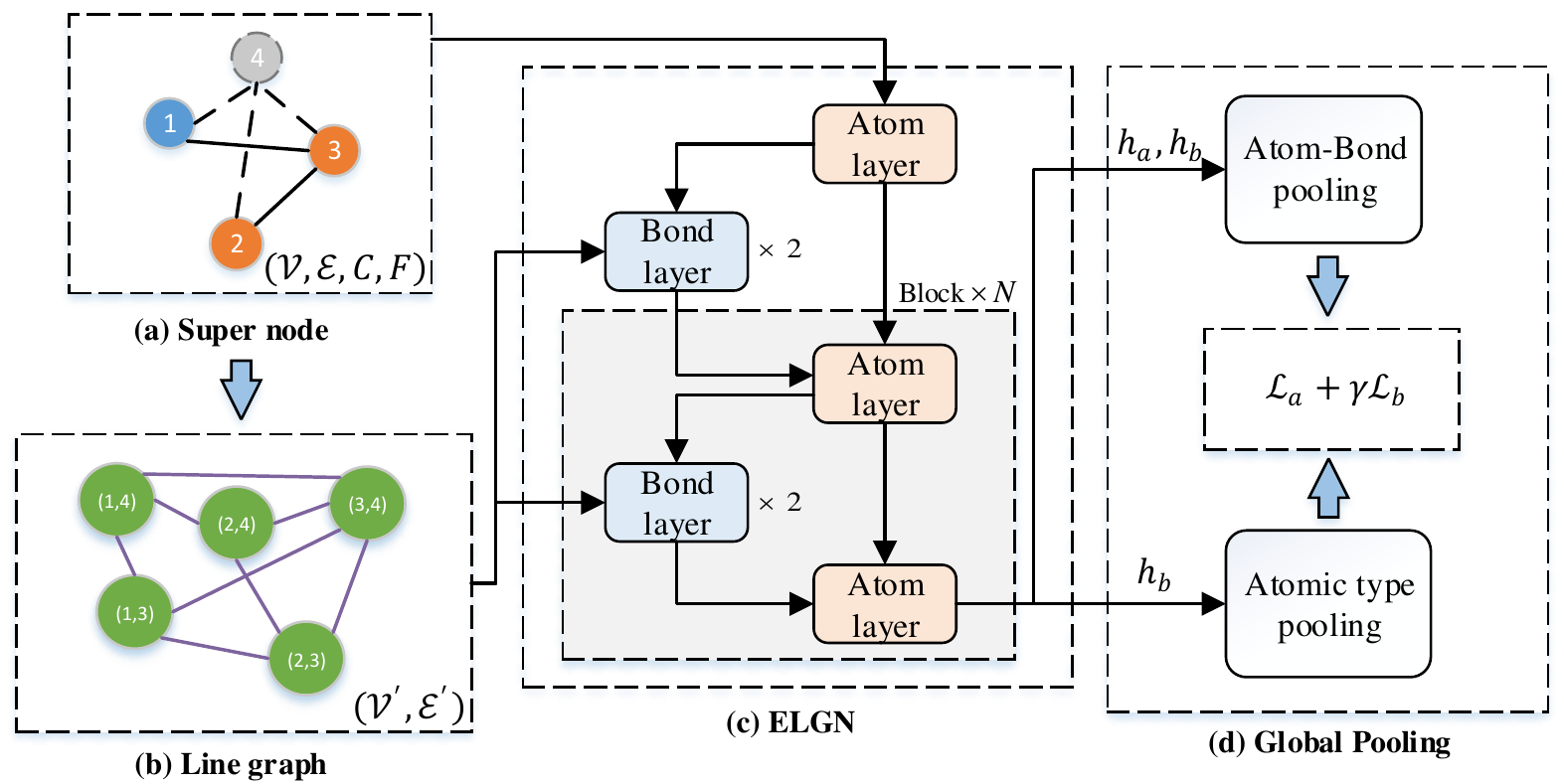} 
\vspace{-0.1in}
\caption{Illustration of equivariant line graph  network (ELGN) model architecture.  (a): The super node, or node 4, forms a "virtual" edge with each of the other nodes in the original graph. (b): We utilize the concept of line graph by converting the raw graph into a line graph to capture the topology of edges. (c): The atom layer and bond layer are the  EGCL  and GCN respectively. (d): Global pooling includes atom and bond features as graph-level feature through the readout function and atomic type pooling, which integrates the long-distance interactions between proteins and ligands.}
\label{fig:framework}
\vspace{-0.2in}
\end{figure*}

\section{4 \quad Method}
\vspace{-0.05in}
In this section, we present our proposed Equivariant Line Graph Network (ELGN) for protein-ligand complex binding affinity prediction. We first introduce the overview of ELGN, and then elaborate on its core components.

\subsection{4.1 \quad Overview}
\vspace{-0.05in}

We present the overall framework of ELGN, as shown in Figure~\ref{fig:framework}. In general, taking the interaction graph $\mathcal{G}$ and interaction line graph $\mathcal{G}'$ as the inputs, ELGN models the 3D spatial structure of the protein-ligand interaction by atom-bond blocks, which generating atom and bond embeddings by neural message passing~\cite{mpnn_icml17}. Then, we use global pooling layers to readout the embeddings for predicting the binding affinity and the co-occurrent frequency of atom pairs as a self-supervised auxiliary task.
Concretely, an atom-bond block is composed of 2 atom layers and 2 bond layers. The atom layer transforms atom and bond embeddings to generate new embeddings for atom and bonds over the protein-ligand interaction graph $\mathcal{G}$, and the bond layer further refine the embeddings of the bonds over the interaction line graph $\mathcal{G}'$. Both the atom layer and the bond layer are GNN layers that consist of the message aggregation and embedding update paradigm, where the output coordinate embeddings are guaranteed to be equivariant with respect to rotation and translations on the initial atom coordinates.
For the global pooling layers, apart from an aggregation pooling to predict the affinity, we also deploy an atom type pooling that predicts the co-occurrent frequency of atom pairs in a larger region of ligand and protein interface as a self-supervised learning task.
The self-supervised task aims to enhancing the atom and bond embeddings, by leveraging the co-occurrence correlations of key atom pairs and capturing long-range interactions of atoms which may not be covered by the interaction graph.
In the following, we will elaborate on the details of atom and bond layers in Section 4.3 and the global pooling layers in Section 4.4.



\subsection{4.2 \quad Line Graph}
\vspace{-0.05in}
The Line graph \cite{article} uses the edges in the original graph as nodes, which alters the viewpoint from which to view the position of edges. Additionally, it is easier to get the edge's topological data. Consequently,  We first transform the interaction graph $\mathcal{G}$ into a line graph $\mathcal{G}'$, and then leverage the line graph to capture the topology of the biomolecular bonds.

\subsection{4.3 \quad Equivariant Line Graph  Network}
\vspace{-0.05in}
3D spatial information is crucial for molecular representation learning. The intuitive method is to learn the embedding of the graph with a 3D coordinate. To this end,  \cite{egnn_icml21} proposes equivariant graph convolutional layer (EGCL) by adding coordinates to message passing and updating coordinates, which benefits in improving the learning of spatial information. In predicting molecular properties, the equivariance is particularly important for proteins and ligands. \cite{chiro_iclr22,3dlinker_icml22} Additionally, the chemical bond has a significant impact on the physical and chemical properties of proteins and ligands \cite{dimenet_iclr20}. Consequently, in our model, we employ the atom layers and bond layers to learn how to represent the 3D protein ligand complex, and the topological structure of the edges.

\subsubsection{Atom Layer.} 
Existing methods treat the protein ligand complex as a 2D graph with node feature extracted using local coordinate system, such as angles between bonds and lengths between nodes, etc~\cite{pgal_kdd21,digeot_iclr22}. 
However, they are unable to capture the global coordinate information, especially, those between an atom and its $k$-hop neighbors (k>4).
To overcome this limitation, we will adopt EGCL as an key component of our ELGN model,
since it can directly learn from the global coordinates, and output equivariant coordinate embeddings with respect to rotations, and translations on the input coordiantes.

Specifically, a EGCL layer is used as an atom layer, which is defined as follows:
\begin{equation}
\h_{a_{i}}^{l}, \h_{b_{i j}}^{l}, \c_{i}^{l}=\mathbf{\varphi }_{a} \left ( \h_{a_{i}}^{l-1}, \h_{a_{j}}^{l-1},\left \| \c_{i}^{l-1} - \c_{j}^{l-1} \right \|^{2}, \h_{b_{i j}}^{l-1}\right )  \label{1}
\end{equation}
where $\varphi_{a}\left(\cdot\right)$ is an non-linear operation.
$\varphi_{a}\left(\cdot\right)$ takes the data from the $l-1$ layer as input, and output three embeddings, $\h_{a_{i}}^{l}$ , $\h_{b_{i j}}^{l}$ and $\c_{i}^{l}$. 
Among them, $\h_{a_{i}}^{l}$ is the $l$-th layer embedding of the atom $a_{i}$. 
$\h_{b_{i j}}^{l}$ is the $l$-th layer embeddding of the bond $b_{ij}$. 
$\c_{i}^{l}$ is the $l$-th layer coordinate embedding of the atom $a_{i}$.

For initializing these embeddings, we set $h_{a_{i}}^{0}=\sigma\left(W F_{a_{i}}\right)$, where $\sigma$ is the RELU function.
If the bond's raw attribute is absent, we initialize $h_{b_{i j}}^{0}=\sigma\left(W \left[F_{a_{i}}||F_{a_{i}}||d_{i j}\right]\right)$ where $d_{i j}=\left \| \textbf{C}_{i}-\textbf{C}_{j} \right \|^{2}$, otherwise, $h_{b_{i j}}^{0}=\sigma \left(W F_{b_{i j}}\right)$. 

Now we will show how to compute the three embeddings: $\h_{a_{i}}^{l}$ , $\h_{b_{i j}}^{l}$ and $\c_{i}^{l}$.
Firstly, the embedding $\h_{b_{i j}}^{l}$ is defined in the Eq. \ref{2}:
\begin{equation}
\h_{b_{i j}}^{l}=\sigma \left ( W_{b}\left [ \h_{a_{i}}^{l}\left |  \right | \h_{a_{j}}^{l} \left |  \right |d_{i j}^{l-1}\left |  \right |  \h^{l-1}_{b_{i j}}\right ]  \right ) \label{2}
\end{equation}
where $\sigma\left(\cdot \right)$ is the activation function and $W_{b}$ is a learnable parameter and $d_{i j}^{l-1}=\| \c_{i}^{l-1}-\c_{j}^{l-1}\|^{2}$. $||$ is the concatenation operation. 
This equation indicates that the bond embedding is updated based on the node information and the bond length.

Secondly, the $i$-th coordinate embedding is updated by the following equation:
\begin{equation}
\mathbf{c}_{i}^{l} =\mathbf{c}_{i}^{l-1}+ \frac{1}{(n-1)} \sum_{j \neq i}\left(\mathbf{c}_{i}^{l-1}-\mathbf{c}_{j}^{l-1}\right) \sigma\left(W_{c}\mathbf{h}_{b_{i j}}^{l}\right) 
\end{equation}
where $W_{c}$ is also a learnable parameter. 
This equation update the coordinate embedding with the help of the bond embedding.
It also use relative coordinates for coordinate embedding updates, which can guarantee the equivariance of network with respect to rotation and translation on the input coordinates.
This way substantially improves the capacity to gather spatial information, which is also the major difference from the existing approaches~\cite{pgal_kdd21}.

After that, the message for node $i$, is computed by aggregating bond embeddings $\h_{b_{i j}}^{l}$ by the following equation: 
\begin{equation}
\mathbf{m}_{a_{i}}^{l} =\sum_{j \neq i} \mathbf{h}_{b_{i j}}^{l} \label{4}
\end{equation}
Finally, the atom embedding of node $i$ is updated using the message $\m_{a_{i}}^{l}$ as follows:
\begin{equation}
\h_{a_{i}}^{l}=\sigma \left ( W_{a}\left [ \h_{a_{i}}^{l-1}|| \m_{a_{i}}^{l}||F_{a_{i}} \right ]  \right ) \label{5}
\end{equation}
where $W_{a}$ is a learnable parameter. 

According to the above formulation, it can be observed that EGCL can directly learn the spatial information of proteins and ligands using global coordinates.
So, it can effectively learn not only the information from 2D graph, such as neighbor knowledge and topology, but also the spatial information of the interaction graph $\mathcal{G}$. 
This makes it very suitable to tasks with 3D structures, such as binding structure prediction~\cite{stark2022equibind}, .

\subsubsection{Bond Layer.} 
The properties of proteins and ligands are significantly influenced by their chemical bonds \cite{dimenet_iclr20}.  
However, the topology between the bonds is neglected by the atom layer, i.e., EGCL.
To this end, a bond layer based on GCN is introduced to learn the topological information of bonds, i.e., the line graph $\mathcal{G}'$.
The bond layer is defined as follows:
\begin{equation}
\h_{b_{i j}}^{l} =\varphi_{b}\left(\mathbf{h}_{b_{k  i}}^{l-1}, \mathbf{h}_{b_{i j}}^{l-1}\right),\quad b_{k i},b_{i j} \in \mathcal{V}' \label{6}
\end{equation}
where $\varphi_{b}\left(\cdot\right)$ is a non-linear operation for the bond layer. 
The input $\h^{l-1}_{b_{i j}}$ is the edge (bond) embedding in the interaction graph $\mathcal{G}$.
The output $\h^l_{b_{i j}}$ is the node embedding in the line graph $\mathcal{G}'$. 
In this way, we only learn the topological information of nodes, and ignore the embedding of edges in the line graph $\mathcal{G}'$.

Specifically, the bond layer consists of the follows three steps:
\begin{equation}
h_{\left(b_{k i},b_{i j}\right)}^{l}= \h_{b_{k i}}^{l-1}, \quad \left(b_{k i}, b_{i j}\right) \in \mathcal{E}' \label{7}
\end{equation}
\begin{equation}
\mathbf{m}_{b_{i j}}^{l} =\sum_{b_{k i} \in \mathcal{N}(b_{i j})} \mathbf{h}_{\left(b_{k i},b_{i j} \right)}^{l}
\end{equation}
\begin{equation}
\mathbf{h}_{b_{i j}}^{l} =\sigma\left(W_{b_{L}}\left[\mathbf{h}_{b_{i j}}^{l-1}|| \mathbf{m}_{b_{i j}}^{l}\right]\right)
\end{equation}
where $W_{b_{L}}$ are two learnable parameters, and $\mathcal{N}(b_{i j})$ is the neighboring nodes of node $b_{i j} \in \mathcal{V}'$.
It further enhanced the model performance to include the topological details of the bond, to the original bond embedding, as later shown in Table 1.

\subsubsection{Atom-Bond Block.} 
As shown in Figure \ref{fig:framework}(c), ELGN consists of a number of atom-bond blocks which contains both atom and bond layers.  
Because it is crucial for the atom-bond layer to learn the spatial information, it is necessary to make the atom-bond layer equivariant~\cite{egnn_icml21}. 
According the equations for the atom-bond block, it is easy to verify that the bond layer merely adds the topological structure of bonds to the bond embedding through GCN and does not update the coordinate embedding or node embedding of the interaction graph $\mathcal{G}$.
As a result, the atom-bond block is equivariant with respect to the rotations and translations on the input coordinate embeddings.

\begin{figure}[]
\centering
\includegraphics[]{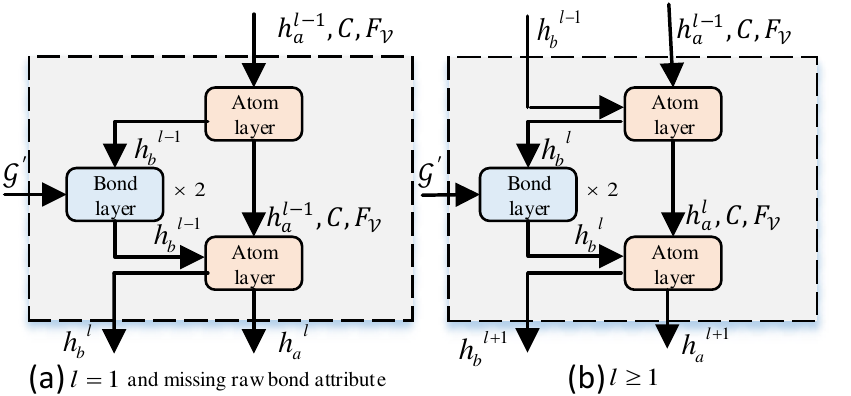} 
\caption{Atom-Bond Block. (a) Only when the raw bond attribute of the complex graph is absent and $l=1$, the atom layer is used to initialize the bond embedding $h_{b_{i j}}^{0}=\sigma\left(W \left[F_{a_{i}}||F_{a_{i}}||d_{i j}\right]\right)$ before the topology information is added. (b) Other times, both the atom and the bond embedding are fed into the atom layer. }
\label{fig2}
\vspace{-0.2in}
\end{figure}

The algorithm flow of the Atom-Bond Block is shown in Figure \ref{fig2}. Firstly, we link the atom layer and the bond layer in parallel.
The atom layer produces the initial bond embedding if the bond’s raw attribute is absent and atom embedding (Figure \ref{2}(a)).
Then the bond layer receives the bond embedding and line graph to learn the bond topology. Finally, the bond embedding, the atom embedding, and the coordinates are fed into the atom layer. 
The model is capable of learning both  topology of bonds and the 3D spatial information of protein-ligand interaction complex.

\subsection{4.4 \quad Global Pooling}
In this subsection, we will introduce two poolings: atom-bond pooling for affinity prediction, and atomic type pooling for predicting an atomic pair approximation matrix.

\subsubsection{Atom-Bond Pooling.}
While the embedding of nodes and edges are suitable to node and edge tasks, 
it is necessary to output graph embedding for graph level tasks, such as graph classification, graph property prediction, etc. 
In fact, the graph embedding usually is computed from the node and edge embedding, by a simple pooling operation, such as sum, mean, max, and min~\cite{mpnn_icml17,velivckovic2019neural}.
Is is obvious that the protein-ligand binding affinity prediction is a graph-level task. 
So, in this paper, the atoms and bonds embedding outputed by ELGN are summed, and then  concatenated as the graph-level embedding, as shown in the following equations:
\begin{equation}
\h_{\mathcal{G}} = \left[\h_{a}||\h_{b}\right]
\end{equation}
\begin{equation}
\h_{a}=\sum_{a_{i}\in \mathcal{V} }\h_{a_{i}}, \quad \h_{b}=\sum_{b_{i j}\in \mathcal{V}' }\h_{b_{i j}}
\end{equation}
where $\h_{\mathcal{G}}$ is the graph-level embedding of the complex graph $\mathcal{G}$, that includes the atom embedding $\h_{a}$, and the bond embedding $\h_{b}$. 

The graph embedding is then fed into a MLP layers to predict protein and ligand binding affinity, which is shown below:
\begin{equation}
\hat{y}=MLP\left( \h_{\mathcal{G}}\right)
\end{equation}
Then , a L1 regression loss is adopted to optimize the model, so that the predicted protein-ligand binding affinity $\hat{y}_{\mathcal{G}}$ is as close as possible to the ground truth $y_{\mathcal{G}}$, which is formulated as the following equation:
\begin{equation}
\label{eq:loneloss}
\mathcal{L}_{a}=\sum_{\mathcal{G} \in \mathcal{D}} \left | \hat{y}_{\mathcal{G}} - y_{\mathcal{G}} \right | 
\end{equation}
where $\mathcal{D}$ is the training set.

\subsubsection{Atomic Type Pooling.} 
The complex graph $\mathcal{G}$ contains all the ligand atoms, and a few protein atoms near the ligand.
However, the protein atoms far away from the ligand also influence the binding affinity. 
To considering their effect, our model is designed to predict the number of interactions for a certain protein-ligand atomic type pair within a given range of distances $d_{\phi}$ \cite{Ballester2010AML}. 
The atom kinds of proteins and ligands are $T_{a}^{P} = \left \{C,N,O,F,P,S,Cl,Br,I\right\}$ and $ T_{a}^{L}=\left \{C,N,O,S\right\}$ for the experimental datasets. For each data, a  atomic pair matrix  $M \in \mathbb{R}^{9 \times 4}$ can be computed as:
\begin{equation}
M_{p l}=\frac{c\left ( \mathcal{T}_{p},\mathcal{T}_{l} \right ) }{ { \sum_{\left ( \mathcal{T}_{i},\mathcal{T}_{j} \right )\in T_{a}^{P} \times T_{a}^{L}  }} c\left ( \mathcal{T}_{i},\mathcal{T}_{j} \right ) } 
\end{equation}
\begin{equation}
c\left ( \mathcal{T}_{p}, \mathcal{T}_{l} \right )=\sum_{a_{i}\in \mathcal{V}_{p}} \sum_{a_{j}\in \mathcal{V}_{l}}  \delta \left ( \tau \left ( a_{i},a_{j} \right ),\left( \mathcal{T}_{p},\mathcal{T}_{l} \right) \right )\Theta \left ( d_{\phi } -d_{i j}\right )  
\end{equation}
where $\mathcal{T}_{p} , \mathcal{T}_{l}$ are atomic numbers of atom types, respectively, e.g., $\mathcal{T}_{l} \in \left\{6,7,8,16\right\}$. The function $\tau \left ( a_{i},a_{j} \right )$ returns the atomic type pair of $(a_{i},a_{j})$, e.g., $\tau \left ( C,N \right )=\left(6,7\right)$. $\delta\left(\cdot,\cdot\right)$ is the Kronecker delta function whose value is 1 if atomic type pair $\tau \left ( a_{i},a_{j} \right )$ is $\left(\mathcal{T}_{p} , \mathcal{T}_{l}\right)$ and 0 otherwise. $\Theta\left(\cdot\right)$ is the Heaviside step function to ensure that the distance between $a_{i}$ and $a_{j}$ is less than $d_{\phi}$. 

To help acquire long-distance information~\cite{pgal_kdd21}, our model tries to predict $M$, using the following equation:
\begin{equation}
\tilde{M}_{p l}=\frac{exp\left ( W h_{p,l} \right ) }{ { \sum_{i,j}exp\left ( W h_{i,j} \right ) } }  
\end{equation}
\begin{equation}
h_{p,l}=\sum_{b_{i j}\in \mathcal{E}}\delta \left ( \tau \left ( a_{i},a_{j} \right ),\left( \mathcal{T}_{p},\mathcal{T}_{l} \right) \right )W_{h}h_{b_{i j}}
\end{equation}
where $W$ and $W_{h}$ are learnable parameters, and $h_{p,l}$ is the atomic type pair embedding which is the bond embeddings of all the same atomic type pairs.
For this prediction, a loss function is given below:
\begin{equation}
\mathcal{L}_{b}=\sum_{\mathcal{G}\in \mathcal{D} }\left \| \tilde{M}   - M  \right \|  
\end{equation}

\subsection{4.5 \quad Optimization Objective}
\vspace{-0.05in}
The ultimate optimization objective is to minimize the total loss $\mathcal{L}$ in Eq.~\ref{eq:finalloss}:
\begin{equation}
\label{eq:finalloss}
\mathcal{L}=\mathcal{L}_{a}+\gamma \mathcal{L}_{b}   
\end{equation}
where $\mathcal{L}_{a}$ is to predict protein-ligand binding affinities, $\mathcal{L}_{b}$ can help to incorporate long-range interactions, and $\gamma$ is a hyper-parameter to tune the influence of long-range distance factors on the prediction of affinity.

\begin{table*}[]
\footnotesize
\caption{Experimental results of our proposed ELGN along with all baselines on PDBbind core set and CSAR-HiQ set.}
\vspace{-0.1in}
\begin{tabular}{@{}l|cccccccc@{}}
\toprule
\multirow{2}{*}{Model} & \multicolumn{4}{c}{PDBbind core set}                                                        & \multicolumn{4}{|c}{CSAR-HiQ set}                                       \\ \cmidrule(l){2-9} 
                       & RMSE $\downarrow$ & MAE $\downarrow$ & SD $\downarrow$ & \multicolumn{1}{c|}{R $\uparrow$}  & RMSE $\downarrow$ & MAE $\downarrow$ & SD $\downarrow$ & R $\uparrow$  \\ \midrule
GGCN                   & 1.735(0.034)      & 1.343 (0.037)    & 1.719 (0.027)   & \multicolumn{1}{c|}{0.613 (0.016)} & 2.324 (0.079)     & 1.732 (0.065)    & 2.302 (0.061)   & 0.464 (0.047) \\
GGAT                   & 1.765 (0.026)     & 1.354 (0.033)    & 1.740 (0.027)   & \multicolumn{1}{c|}{0.601 (0.016)} & 2.213 (0.053)     & 1.651 (0.061)    & 2.215 (0.050)   & 0.524 (0.032) \\
GGIN                   & 1.640 (0.044)     & 1.261 (0.044)    & 1.621 (0.036)   & \multicolumn{1}{c|}{0.667 (0.018)} & 2.158 (0.074)     & 1.624 (0.058)    & 2.156 (0.088)   & 0.558 (0.047) \\
GGACN              & 1.562 (0.022)     & 1.191 (0.016)    & 1.558 (0.018)   & \multicolumn{1}{c|}{0.697 (0.008)} & 1.980 (0.055)     & 1.493 (0.046)    & 1.969 (0.057)   & 0.653 (0.026) \\ \midrule
DMPNN                  & 1.493 (0.016)     & 1.188 (0.009)    & 1.489 (0.014)   & \multicolumn{1}{c|}{0.729 (0.006)} & 1.886 (0.026)     & 1.488 (0.054)    & 1.865 (0.035)   & 0.697 (0.013) \\
CMPNN                  & 1.408 (0.028)     & 1.117 (0.031)    & 1.399 (0.025)   & \multicolumn{1}{c|}{0.765 (0.009)} & 1.839 (0.096)     & 1.411 (0.064)    & 1.767 (0.103)   & 0.730 (0.052) \\ \midrule
SGCN                   & 1.583 (0.033)     & 1.250 (0.036)    & 1.582 (0.320)   & \multicolumn{1}{c|}{0.686 (0.015)} & 1.902 (0.063)     & 1.472 (0.067)    & 1.891 (0.077)   & 0.686 (0.030) \\
GNN-DTI                & 1.492 (0.025)     & 1.192 (0.032)    & 1.471 (0.051)   & \multicolumn{1}{c|}{0.736 (0.021)} & 1.972 (0.061)     & 1.547 (0.058)    & 1.834 (0.090)   & 0.709 (0.035) \\
MAT                    & 1.457 (0.037)     & 1.154 (0.037)    & 1.445 (0.033)   & \multicolumn{1}{c|}{0.747 (0.013)} & 1.879 (0.065)     & 1.435 (0.058)    & 1.816 (0.083)   & 0.715 (0.030) \\
DimeNet                & 1.453 (0.027)     & 1.138 (0.026)    & 1.434 (0.023)   & \multicolumn{1}{c|}{0.752 (0.010)} & 1.805 (0.036)     & 1.338 (0.026)    & 1.798 (0.027)   & 0.723 (0.010) \\
SIGN                   & 1.316 (0.031)     & 1.027 (0.025)    & 1.312 (0.035)   & \multicolumn{1}{c|}{0.797 (0.012)} & 1.735 (0.031)     & 1.327 (0.040)    & 1.709 (0.044)   & 0.754 (0.014) \\ \midrule
ELGN                   & \textbf{1.285(0.027)}      & \textbf{1.013(0.022)}     & \textbf{1.263(0.026)}    & \textbf{0.810(0.012)}                       &\multicolumn{1}{|c}{\textbf{1.628(0.026)}}    & \textbf{1.260(0.025)}     & \textbf{1.594(0.022)}    & \textbf{0.762(0.017)}  \\ \bottomrule
\end{tabular}
\vspace{-0.1in}
\end{table*}

\section{5 \quad Experiments}

\subsection{5.1 \quad Experiment Settings}
\subsubsection{Dataset.} The datasets PDBbind \cite{wang2005pdbbind} and CSAR-HiQ \cite{dunbar2011csar}  are used to assess our model for predicting protein-ligand binding affinity.  In addition, these datasets are used to evaluate the accuracy of various scoring functions for protein-ligand binding affinities. PDBbind contains experimentally determined binding affinities for protein ligand complexes and has different versions. In our work, we use PDBbind v2016 which includes general, refined and core set. For these subsets, overlapping data is present. There are 13283 complexes in the general set, and 4057 complexes from the refined set are chosen because they are of higher quality. A higher-quality refined set is chosen to make up the core set which contains 290 complexs. CSAR-HiQ is an additional dataset used to assess the generalizability of the model. It has two subsets with 176 and 167 protein-ligand complexs.
\subsubsection{Evaluation Metrics.} In our work, we use four metrics \cite{bty374,pgal_kdd21} to evaluate the performance of the models, including Root Mean Square Error (RMSE) and Mean Absolute Error (MAE), Pearson correlation coefficient (R) and the standard deviation (SD) in regression. 
Pearson correlation coefficient measures the correlation between  the predictions and the ground truth values.
\subsubsection{Baselines.} 
We compare our ELGN with state-of-the-art baselines from 3 categories as follows:
\begin{itemize}
    \item \textbf{GraphDTA methods.} GraphDTA \cite{nguyen2021graphdta} utilizes CNN to encode a protein sequence and GNN to encode a ligand graph, respectively. 
    We compare with four model variants equipped with different GNN layers, i.e., GCN (GGCN) \cite{gcn_iclr17}, GAT (GGAT) \cite{gat_iclr18}, GIN (GGIN) \cite{gin_iclr19} and a combined GAT-GCN (GGACN).

    \item \textbf{Message passing-based methods.} DMPNN \cite{yang2019analyzing} is an edge-based message passing neural network and  utilizes the edges' aggregation procedure to learn the spatial information between atoms.  CMPNN \cite{song2020communicative} enhances the communication between nodes and edges through a communicative kernel to improve the molecular embedding.
    
    \item \textbf{Geometry learning-based GNN methods.} GNN-DTI \cite{lim2019predicting} is a distance-aware graph attention algorithm to differentiate various types of intermolecular interactions. SGCN \cite{danel2020spatial} leverages node positions based on GCN.  MAT \cite{maziarka2020molecule} will use molecular graph structure and inter-atomic distances to improve the attention mechanism in Transformer. DimeNet \cite{dimenet_iclr20} uses the directional message passing framework and employs the Bessel functions to encode the angle and distance information. SIGN \cite{pgal_kdd21} leveraging the fine-grained structure and long-range interaction information among protein and ligand to improve the complex embedding.

\end{itemize}


\subsubsection{Implementation Details.}
Construction of the initial features refers to previous work~\cite{bty374,pgal_kdd21}, where an atom possesses 18-dim raw features containing atomic types, hybridization, atom properties, etc.
An atom is encoded by 37-dim vector where the $1$-$18$-th dimensions are for a ligand atom and the $19$-$36$-th dimensions are for a protein atom. For a ligand atom, the $19$-$36$-th dimensions are all zeros and vice versa. In addition, the 37-th dimension is a binary indicator for the super node. To construct the interaction graph and the atomic type  matrix, we set $d = 5 \AA$ and $d_{\phi}=12 \AA$ following~\cite{muegge1999general}. 

For testing PDBbind core set and CSAR-HiQ, we construct two ELGN models with 2 and 1 atom-bond blocks, and train the two models on PDBbind refined set by 400 and 50 epochs, respectively.
We use Adam optimizer with a learning rate of 0.001 and a batch size of 32. A dropout rate of 0.2 is used to prevent overfitting. The hyper-parameter $\gamma$ in Eq.~\ref{eq:finalloss} is set to 1.75. 
The final parameters of the model is the average of the parameters of the top 5 model checkpoints on the evaluation set during training. 
All the experiments are conducted on one Tesla V100 GPU and Inter Xeon Gold 5218 16-Core Processor. And the performance of all the baselines refers to  \cite{pgal_kdd21}. 

\subsection{5.2 \quad Performance Evaluation}

\subsubsection{Overall Comparison.} 
Table 1 lists the performance of ELGN compared with the baselines in the 3 categories on PDBbind v2016 core set and CSAR-HiQ, where the results are the average and the standard deviation of five random runs. 
In general, the ELGN outperforms all the baselines and improves the performance of RMSE over the best baselines with 2.3\% and 6.2\%, respectively. 

The 4 GraphDTA baselines learn the embedding by the sequence of the protein, they ignore the spatial structure of the protein, which leads to their poor performance. In order to learn spatial information, SGCN, GNN-DTI and MAT use position and distance information, DimeNet learns the angle information, they can learn partial spatial information, and the performance is improved. In addition, CMPNN uses distance information to enhance message passing in DMPNN and achieves better performance. SIGN takes into account the interaction between protein and ligand, and learns angle information through polar coordinates, which greatly improves the effect, but 3D spatial information still cannot be fully learned. We use ELGN to learn the equivariant spatial information based on the 3D protein ligand complex and consider the topological information of the bond. Therefore, The ELGN can learn the complex spatial information well and achieve the best performance.

\begin{figure}[h]
\centering
\includegraphics[width=1.0\linewidth]{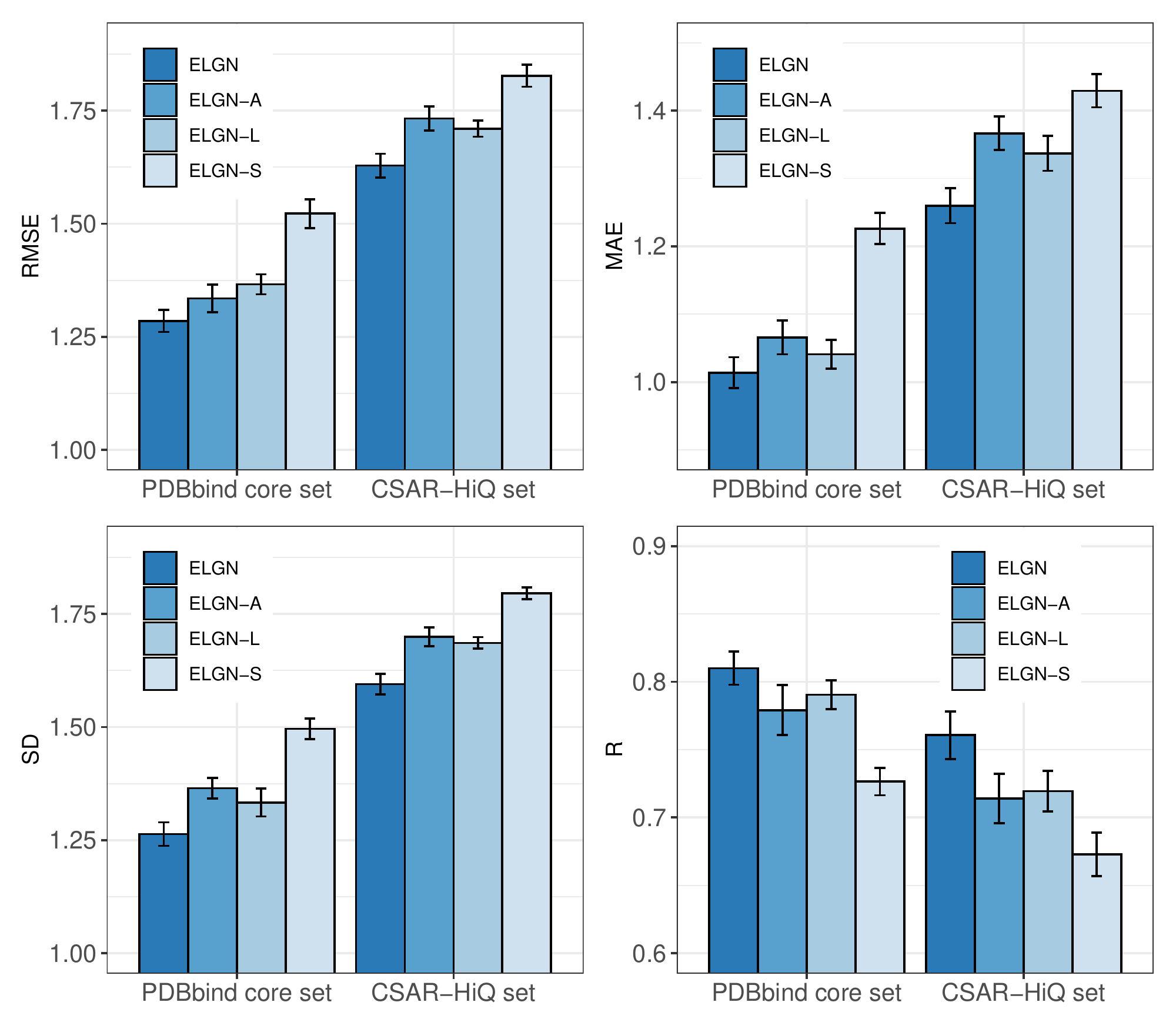} 
\vspace{-0.2in}
\caption{The ablated variants of ELGN}
\label{fig3}
\vspace{-0.2in}
\end{figure}

\subsubsection{Ablation Studies.} 
To investigate the effectiveness of the key components of ELGN, i.e., the atomic type pooling, modeling on the line graph and the super node, we compare with three variants of ELGN as ablation studies in the following:
(1) \textbf{ELGN-A} removes the atomic type pooling and computes the loss only by Eq.~\ref{eq:loneloss}.
(2) \textbf{ELGN-L} removes bond layers for the line graph, thereby all the embeddings as well as the coordinates are only updated by the atom layers.
(3) \textbf{ELGN-S} removes the super node from the protein-ligand interaction graph, where atom layers operate on the original interaction graph. 

Figure \ref{fig3} shows the comparison of metrics for three ablation studies on the two benchmark datasets. Our proposed ELGN outperforms three variants, demonstrating that ELGN can learn 3D spatial structure well for the prediction binding affinity of protein-ligand complex. Additionally, ELGN-S demonstrates the global information of protein ligand complex  contributes to the affinity prediction. And the performance of ELGN-L and ELGN-A is worse than ELGN, proving that the topological information of bonds and long-range interactions between proteins and ligands are beneficial for 3D spatial information learning.
 
\section{6 \quad Conclusion}
\vspace{-0.05in}
This paper aims to improve affinity prediction by learning the spatial information of 3D protein-ligand complexes. We propose a new equivariant line graph network (ELGN) to learn representations of 3D protein-ligand complexes by leveraging the local and global spatial information. We employ a super node and an atomic type pooling to capture global information between proteins and ligands. Then we design an equivariant atom-bond block to learn local information of protein-ligand complexes and leverage a line graph for topology of bonds. The experiments show that ELGN surpasses previous methods with better effectiveness and generalizability.

\nobibliography{aaai23}

\section{Reference}
\bibentry{pgal_kdd21}.\\[.2em]
\bibentry{egnn_icml21}.\\[.2em]
\bibentry{gcn_iclr17}.\\[.2em]
\bibentry{article}.\\[.2em]
\bibentry{bandyopadhyay2019beyond}.\\[.2em]
\bibentry{li2017learning}.\\[.2em]
\bibentry{dn4gl_icml22}.\\[.2em]
\bibentry{Ballester2010AML}.\\[.2em]
\bibentry{digeot_iclr22}.\\[.2em]
\bibentry{dimenet_iclr20}.\\[.2em]
\bibentry{mpnn_icml17}.\\[.2em]
\bibentry{velivckovic2019neural}.\\[.2em]
\bibentry{nt2019revisiting}.\\[.2em]
\bibentry{expgnn_iclr20}.\\[.2em]
\bibentry{bty374}.\\[.2em]
\bibentry{dunbar2011csar}.\\[.2em]
\bibentry{nguyen2021graphdta}.\\[.2em]
\bibentry{gat_iclr18}.\\[.2em]
\bibentry{gin_iclr19}.\\[.2em]
\bibentry{danel2020spatial}.\\[.2em]
\bibentry{lim2019predicting}.\\[.2em]
\bibentry{yang2019analyzing}.\\[.2em]
\bibentry{maziarka2020molecule}.\\[.2em]
\bibentry{song2020communicative}\\[.2em]
\bibentry{muegge1999general}\\[.2em]
\bibentry{bahuguna2020overview}\\[.2em]
\bibentry{theodoris2021network}\\[.2em]
\bibentry{callaway2020will}\\[.2em]
\bibentry{edm_icml22}\\[.2em]
\bibentry{pocket2mol_icml22}\\[.2em]
\bibentry{graphbp_icml22}\\[.2em]
\bibentry{shan2022deep}\\[.2em]
\bibentry{3dlinker_icml22}\\[.2em]
\bibentry{chiro_iclr22}\\[.2em]
\bibentry{stark2022equibind}\\[.2em]
\bibentry{ozturk2018deepdta}\\[.2em]
\bibentry{ragoza2017protein}\\[.2em]
\bibentry{zheng2019onionnet}\\[.2em]
\bibentry{thafar2022affinity2vec}\\[.2em]
\bibentry{decgcn_neurips21}\\[.2em]
\bibentry{gnnmp_iclr20}\\[.2em]
\bibentry{li2022bacpi}\\[.2em]
\bibentry{plgnn_aaai22}\\[.2em]
\bibentry{se3gnn_icml22}\\[.2em]
\bibentry{nacher2004clustering}\\[.2em]
\bibentry{pereira2004detection}\\[.2em]
\bibentry{liu2022graph}\\[.2em]
\bibentry{hsu2022efficient}\\[.2em]

\newpage
\appendix
\section{Appendix}
\section{A Details of ELGN}
\subsection{A.1 ELGN}
We give the simplified pseudocode for our propose ELGN in Algorithm \ref{elgn:algorithm}.

\begin{algorithm}[]
\caption{ELGN}
\label{elgn:algorithm}
\textbf{Input}: The protein-ligand interaction graph $\mathcal{G}$ and the interaction line graph $\mathcal{G}'$ \\
\textbf{Parameter}: The  number $N$ of Atom-bond bolck\\
\textbf{Output}: $\textbf{h}_{a}$, $\textbf{h}_{b}$, $\textbf{c}$
\begin{algorithmic}[1] 
\STATE Initialize $l \leftarrow 1, n \leftarrow 1$
\IF {$\textbf{F}_{b}$ is none}
\STATE $\textbf{h}_{b_{i j}}^{0} =$ Linear(concat($\textbf{F}_{a_{i}},\textbf{F}_{a_{j}},d_{i j}$))
\ELSE 
\STATE $\textbf{h}_{b_{i j}}^{0} = $ Linear($\textbf{F}_{b_{i j}}$)
\ENDIF
\STATE $\textbf{h}_{a}^{0} =$ Linear( $\textbf{F}_{a}$), $\textbf{c}^{0} \leftarrow \textbf{C}$
\WHILE{$n \leq N$}
\STATE $\textbf{h}_{a}^{l+1}$,$\textbf{h}_{b}^{l+1}$, $\textbf{c}^{l+1}= $  AtomBondBlock($\textbf{h}_{a}^{l-1}$,$\textbf{h}_{b}^{l-1}$, $\textbf{c}^{l-1}$)
\STATE $n \leftarrow n + 1$
\STATE $l \leftarrow l + 2$
\ENDWHILE
\STATE $\textbf{h}_{a} \leftarrow \textbf{h}_{a}^{l+1}$, $\textbf{h}_{b} \leftarrow \textbf{h}_{b}^{l+1}$, $\textbf{c} \leftarrow \textbf{c}^{l+1}$
\STATE \textbf{return} $\textbf{h}_{a}$, $\textbf{h}_{b}$, $\textbf{c}$ \\
\textbf{def} AtomBondBlock($\textbf{h}_{a}^{l-1}$,$\textbf{h}_{b}^{l-1}$, $\textbf{c}^{l-1}$):
\STATE \quad $\textbf{h}_{a}^{l}$, $\textbf{h}_{b}^{l}$, $\textbf{c}^{l} = $ AtomLayer($\textbf{h}_{a}^{l-1}$, $\textbf{h}_{b}^{l-1}$, $\textbf{c}^{l-1}$)
\STATE \quad $\textbf{h}_{b}^{l} = $  BondLayer($\mathcal{G}'$,$\textbf{h}_{b}^{l}$)
\STATE \quad $\textbf{h}_{b}^{l} = $  BondLayer$1$($\mathcal{G}'$,$\textbf{h}_{b}^{l}$)
\STATE \quad $\textbf{h}_{a}^{l+1}$,$\textbf{h}_{b}^{l+1}$, $\textbf{c}^{l+1} = $ AtomLayer$1$($\textbf{h}_{a}^{l}$, $\textbf{h}_{b}^{l}$, $\textbf{c}^{l}$)
\STATE \quad \textbf{return} $\textbf{h}_{a}^{l+1}$, $\textbf{h}_{b}^{l+1}$, $\textbf{c}^{l+1}$
\end{algorithmic}
\end{algorithm}
\subsection{A.2 Atom Layer}
The atom layer is the key component in the ELGN. And we give  the pseudocode of the atom layer in Algorithm \ref{elgn:algorithm1}.
\begin{algorithm}[]
\caption{Atom Layer}
\label{elgn:algorithm1}
\textbf{def} AtomLayer($\mathcal{G},\textbf{h}_{a_{i}}^{l-1},\textbf{h}_{a_{j}}^{l-1}, \textbf{h}_{b_{i j}}^{l-1},\textbf{c}^{l-1}, \textbf{F}_{a_{i}}$):

\begin{algorithmic}[1] 
\STATE $\textbf{h}_{a_{i}}^{l-1}, \textbf{h}_{a_{j}}^{l-1} \leftarrow $ Dropout($\textbf{h}_{a_{i}}^{l-1}, \textbf{h}_{a_{j}}^{l-1}$)
\STATE $\textbf{F}_{a_{i}} \leftarrow$ Dropout($\textbf{F}_{a_{i}}$)
\STATE $d_{i j}^{l-1} = \left \| \textbf{c}_{i}^{l-1} - \textbf{c}_{j}^{l-1} \right \|^{2} $
\STATE $\textbf{h}_{b_{i j}}^{l}$ = Linear(concat($\textbf{h}_{a_{i}}^{l-1},\textbf{h}_{a_{j}}^{l-1},d_{i j}, \textbf{h}_{b_{i j}}^{l-1}$))
\STATE $\textbf{h}_{b_{i j}}^{l} \leftarrow$ Dropout($\textbf{h}_{b_{i j}}^{l}$)
\STATE $\textbf{c}_{i}^{l} = \textbf{c}_{i}^{l-1} + \dfrac{1}{n-1}\sum_{j \ne i}(\textbf{c}_{i}^{l-1}-\textbf{c}_{j}^{l-1}) \odot$ Linear($\textbf{h}_{b_{i j}}^{l}$)
\STATE $\textbf{m}_{a_{i}}^{l}=$ Aggregate($\mathcal{G}, \textbf{h}_{b_{i j}}^{l})$
\STATE $\textbf{h}_{a_{i}}^{l}$ = Linear(concat($\textbf{h}_{a_{i}}^{l-1}, \textbf{m}_{a_{i}}^{l}, \textbf{F}_{a_{i}}$))
\STATE \textbf{return}  $\textbf{h}_{a_{i}}^{l}$, $\textbf{h}_{b_{i j}}^{l}$, $\textbf{c}_{i}^{l}$
\end{algorithmic}
\end{algorithm}

\subsection{A.3 Bond Layer}
The bond layer learn the topological information of bonds with interaction line graph in the ELGN. And we give the pseudocode of the bond layer in Algorithm \ref{elgn:algorithm2}.
\begin{algorithm}[]
\caption{Bond Layer}
\label{elgn:algorithm2}
\textbf{def} BondLayer($\mathcal{G}', \textbf{h}_{b_{k i }}^{l-1}, \textbf{h}_{b_{i j}}^{l-1}$):
\begin{algorithmic}[1] 
\STATE $\textbf{h}_{b_{k i }}^{l-1}, \textbf{h}_{b_{i j}}^{l-1} \leftarrow$  Normalize($\textbf{h}_{b_{k i }}^{l-1}, \textbf{h}_{b_{i j}}^{l-1}$)
\STATE $\textbf{h}_{b_{k i }}^{l-1}, \textbf{h}_{b_{i j}}^{l-1} \leftarrow$ Dropout($\textbf{h}_{b_{k i }}^{l-1}, \textbf{h}_{b_{i j}}^{l-1}$)
\STATE $\h_{(b_{k i},b_{i j})}^{l}$ = $\h_{b_{k i}}^{l-1}$
\STATE $\textbf{m}_{b_{i j}}^{l}$ = Aggregate($\mathcal{G}', \textbf{h}_{(b_{k i},b_{i j})}^{l}$)
\STATE $\textbf{h}_{b_{i j}}^{l} = $Linear(concat($\textbf{h}_{b_{i j }}^{l-1}, \textbf{m}_{b_{i j}}^{l}$))
\STATE \textbf{return} $\textbf{h}_{b_{i j}}^{l}$
\end{algorithmic}
\end{algorithm}

\section{B Features}
In the protein-ligand interaction $\mathcal{G}$, we use a totally of 37 features to describe an atom $F_{a}$ according to \cite{bty374} where the $1$-$18$-th dimensions are for a ligand atom and the $19$-$36$-th dimensions are for a protein atom. For a ligand atom, the $19$-$36$-th dimensions are all zeros and vice versa. In addition, the 37-th dimension is a binary indicator for the super node. The 18 features are as follows:
\begin{itemize}
    \item 9 bits (one-hot or all null) encoding atom types: B, C, N, O, P, S,Se, halogen and metal
    \item 1 integer (1, 2, or 3) with atom hybridization: hyb
    \item 1 integer counting the numbers of bonds with other heavyatoms:
heavy\_valence
    \item 1 integer counting the numbers of bonds with other heteroatoms:
hetero\_valence
    \item 5 bits (1 if present) encoding properties defined with SMARTS
patterns: hydrophobic, aromatic, acceptor, donor and ring
    \item 1 float with partial charge: partial charge
\end{itemize}

\section{C Experiment Details}

\subsection{C.1 ELGN}
We use the PDBbind v2016 refined set to train two ELGN models with different epochs and  test on the PDBbind core set and CSAR-HiQ set. 

Due to the PDBbind refined and core set having overlapping cases, we delete the cases. We use 3767 protein-ligand complexes in the refined set by randomly splitting with a ratio of 9:1 to train and validate. We remove overlapping  data between the refined set and CSAR-HiQ set, the remaining part has 133 protein-ligand complexes in the CSAR-HiQ set for testing. 
 
 The hidden dimensions of atom and bond representations are set to 128.  

\end{document}